\documentclass{aa}
\usepackage{graphicx}
\usepackage{textcomp}
\usepackage[T1]{fontenc}
\usepackage{times}

\begin{document}

\newcommand{\Msun}{\ensuremath{\mathrm{M}_\odot}}
\newcommand{\Minit}{\ensuremath{\mathrm{M}_{\rm WD, init}}}
\newcommand{\Lsun}{\ensuremath{\mathrm{L}_\odot}}
\newcommand{\Rsun}{\ensuremath{\mathrm{R}_\odot}}
\newcommand{\Mdot}{\ensuremath{\dot{M}}}
\newcommand{\density}{\ensuremath{\mathrm{g/cm^3}}}
\newcommand{\arrow}{$\longrightarrow$}
\newcommand{\lsim}{\mathrel{\hbox{\rlap{\lower.55ex \hbox {$\sim$}}
 \kern-.3em \raise.4ex \hbox{$<$}}}}
\newcommand{\gsim}{\mathrel{\hbox{\rlap{\lower.55ex \hbox {$\sim$}}
 \kern-.3em \raise.4ex \hbox{$>$}}}}
\newcommand{\half}{\ensuremath{{\textstyle\frac{1}{2}}}}
\newcommand{\e}{\ensuremath{\mathrm{e}}}
\newcommand{\n}{\ensuremath{\mathrm{n}}}
\newcommand{\p}{\ensuremath{\mathrm{p}}}
\renewcommand{\H}[1]{\ensuremath{{^{#1}\mathrm{H}}}}
\newcommand{\D}{\ensuremath{\mathrm{D}}}
\newcommand{\He}[1]{\ensuremath{{^{#1}\mathrm{He}}}}
\newcommand{\Li}[1]{\ensuremath{{^{#1}\mathrm{Li}}}}
\newcommand{\Be}[1]{\ensuremath{{^{#1}\mathrm{Be}}}}
\newcommand{\B}[1]{\ensuremath{{^{#1}\mathrm{B}}}}
\newcommand{\C}[1]{\ensuremath{{^{#1}\mathrm{C}}}}
\newcommand{\N}[1]{\ensuremath{{^{#1}\mathrm{N}}}}
\renewcommand{\O}[1]{\ensuremath{{^{#1}\mathrm{O}}}}
\newcommand{\F}[1]{\ensuremath{{^{#1}\mathrm{F}}}}
\newcommand{\Na}[1]{\ensuremath{{^{#1}\mathrm{Na}}}}
\newcommand{\Ne}[1]{\ensuremath{{^{#1}\mathrm{Ne}}}}
\newcommand{\Mg}[1]{\ensuremath{{^{#1}\mathrm{Mg}}}}
\newcommand{\Fe}[1]{\ensuremath{{^{#1}\mathrm{Fe}}}}
\newcommand{\Ni}[1]{\ensuremath{{^{#1}\mathrm{Ni}}}}
\newcommand{\Co}[1]{\ensuremath{{^{#1}\mathrm{Co}}}}
\newcommand{\Al}[1]{\ensuremath{{^{#1}\mathrm{Al}}}}
\newcommand{\avg}[1]{\ensuremath{\langle#1\rangle}}
\newcommand{\rate}[1]{\ensuremath{r_\mathrm{#1}}}
\newcommand{\sv}[1]{\ensuremath{\langle\sigma v\rangle_\mathrm{#1}}}
\newcommand{\life}[2]{\ensuremath{\tau_\mathrm{#1}(\mathrm{#2})}}
\newcommand{\Mwd}{$M_{\rm{WD}}$}
\newcommand{\msyr}{$\rm{M}_{\odot}/\rm{yr}$}
\newcommand{\kmpersec}{$\rm{km}\rm{s}^{-1}$}
\newcommand{\ctw}{$^{12}\rm{C}$}
\newcommand{\ost}{$^{16}\rm{O}$}
\newcommand{\rtrip}{$3\alpha$}
\newcommand{\rche}{$^{12}\rm{C}(\alpha,\gamma)^{16}\rm{O}$}
\newcommand{\mdota}{\ensuremath{2\times10^{-8}}}
\newcommand{\mdotb}{\ensuremath{3\times10^{-8}}}
\newcommand{\DMhe}{\ensuremath{\Delta M_{\rm He}}}
\newcommand{\Mhes}{\ensuremath{M_{\rm HeS}}}

\authorrunning{S.-C. Yoon \& N.  Langer}
\titlerunning{Helium accreting CO white dwarfs with rotation}

\title{Helium accreting CO white dwarfs with rotation: helium novae instead of double detonation}

\author{S.-C. Yoon \and N. Langer}

\institute{Astronomical Institute, Utrecht University, Princetonplein 5,
NL-3584 CC, Utrecht, The Netherlands}

\offprints {S.-C. Yoon, \email{S.Ch.Yoon@astro.uu.nl}}
\date{Received  / Accepted }


\abstract{We present evolutionary models of
 helium accreting carbon-oxygen white dwarfs
 in which we include the effects of 
 the spin-up of the accreting star induced by angular momentum 
 accretion, 
 rotationally induced chemical mixing and rotational energy dissipation.
 Initial masses of 0.6 \Msun{} and 0.8 \Msun{} and constant accretion rates 
 of a few times $10^{-8}{\rm M_{\odot}/yr}$ of helium rich matter
 have been  considered, which is typical for the
 sub-Chandrasekhar mass progenitor scenario for Type Ia supernovae. 
 It is found that the helium envelope in an accreting white dwarf 
 is heated efficiently by friction
 in the differentially rotating spun-up layers. As a result, helium ignites
 much earlier and
 under much less degenerate conditions compared to the corresponding non-rotating case. Consequently, a
 helium detonation may be avoided, which questions the sub-Chandrasekhar
 mass progenitor scenario for Type Ia supernovae. 
 We discuss implications of our results for the 
 evolution of helium star plus white dwarf
 binary systems as possible progenitors of recurrent helium novae.
\keywords{stars: evolution -- stars: white dwarf -- stars: rotation -- stars:novae -- supernovae: Type Ia}
}
\maketitle


\section{Introduction}

Type Ia supernovae are of key importance for the chemical evolution
of galaxies, as they are a major producer of iron group elements 
(e.g. Nomoto et al.~\cite{Nomoto84}; Renzini~\cite{Renzini99} ). 
They were also found to be excellent distance indicator 
and have become an indispensable tool in cosmology 
(Phillips~\cite{Phillips93}; Hamuy et al.~\cite{Hamuy96}; Branch~\cite{Branch98}).
The recent suggestion of a non-zero cosmological constant
is partly based on observations of SNe Ia at high redshift (Leibundgut~\cite{Leibundgut01}).
Given that distance determinations at high redshift through SNe~Ia
depend on the assumption 
of the homogeneity of SNe Ia light curves throughout
the ages, an
understanding of the possible diversity of their
progenitors is crucial to evaluate this approach. 
Nevertheless, the progenitors of Type Ia supernovae have not been identified
yet, and the debate on their exact nature continues (e.g., Livio~\cite{Livio01}).

One possibility to obtain a SN~Ia is the detonation of
the degenerate helium layer accumulated on top of
a CO white dwarf due to mass transfer from its low mass helium star companion
in a close binary system, which triggers a carbon
detonation in the white dwarf core.
This is the so called double detonation or sub-Chandrasekhar mass
scenario for SNe~Ia, as it may allow to explode white dwarfs with masses
well below the Chandrasekhar mass
(e.g. Nomoto~\cite{Nomoto82b}; Fujimoto~\cite{Fujimoto82c};
Limongi \& Tornamb\'e~\cite{Limongi91}; Livne~\cite{Livne90}; 
Woosley \& Weaver~\cite{Woosley94}; Livne \& Arnett~\cite{Livne95}).
While the capability of the helium detonation to ignite the CO core
is still debated (e.g., Livio~\cite{Livio01}), the helium detonation
by itself would produce an explosion of supernova scale. 

Currently, the sub-Chandrasekhar mass scenario is not favored 
as a major source of SNe Ia mainly because
the light curves and spectra obtained from this model
are not in good agreement with observations 
(e.g. H\"oflich \& Khokhlov~\cite{Hoeflich96}; Nugent et al.~\cite{Nugent97}).
Especially, the predicted presence of high velocity Ni and He
is most stringently criticized 
(e.g. Hillebrandt \& Niemeyer~\cite{Hillebrandt00}; Livio~\cite{Livio01}).

On the other hand, 
stellar and binary evolution theory predicts a realization frequency
of binary systems such as helium star cataclysmics ---
which might produce double detonating sub-Chandrasekhar mass white dwarfs 
--- which amounts to a few times $10^{-3}~{\rm yr^{-1}}$ per galaxy
(e.g. Iben \& Tutukov~\cite{Iben91}; Reg\"os et al.~\cite{Regos02})
which is comparable to the expected total SN~Ia rate in the Milky Way.
This raises the question why such explosions are practically never
observed.

We note that the sub-Chandrasekhar mass SN progenitor models 
which have been constructed so far 
neglected the effects of rotation, which
can be one of the primary factors 
determining the evolution of stars, in particular of massive stars
(Langer~\cite{Langer98};  Maeder \& Meynet~\cite{Maeder00}).
Iben \& Tutukov (\cite{Iben91}) pointed out that rotation
may indeed be important in helium star cataclysmic systems.
Yoon \& Langer (\cite{Yoon02}, \cite{Yoon04a}) and Yoon et al. (\cite{Yoon04d}) showed
that effects of rotation might be essential
for the evolution of accreting white dwarfs when
the accreted matter contains a high specific angular momentum. 
The induced spin-up was found to change the white dwarf structure 
significantly and to produce rotationally induced chemical mixing.

In this paper, we suggest that rotation could play a key role
in helium accreting white dwarfs such that in model which would
produce a helium detonation this phenomenon is completely avoided
when the white dwarf spin-up is considered.
After explaining the numerical method and 
physical assumptions of the present study
in Sect.~\ref{sect:method}, we investigate 
the evolution of helium accreting carbon-oxygen white dwarfs
with accretion rates of $\sim 10^{-8}$~\msyr{}, 
with the effects of rotation considered, 
in Sect.~\ref{sect:results}. 
Implications of our results for helium novae and
neutron capture nucleosynthesis are discussed in Sect.~\ref{sect:discussion}. 
Our main conclusions are summarized in Sect.~\ref{sect:conclusion}

\section{Numerical method and physical assumptions}\label{sect:method}

We have computed the numerical models with a hydrodynamic
stellar evolution code (Langer et al.~\cite{Langer88}),
which incorporates the effect of the centrifugal force on the stellar structure
and rotationally induced transport of angular momentum and chemical
species due to the dynamical and secular shear instability, 
the Goldreich-Schubert-Fricke instability and the Eddington-Sweet circulations
(Heger et al.~\cite{Heger00a}; Yoon \& Langer~\cite{Yoon04a}). 
Conservation of angular momentum and energy of viscous fluids
requires dissipation of rotational energy as
angular momentum is transported by viscous friction in 
differentially rotating layers
(e.g. Landau \& Lifshitz~\cite{Landau84}).
This effect is considered in our calculations 
following Mochkovitch \& Livio (\cite{Mochkovitch89}) as:
\begin{equation}\label{eq1}
\epsilon_{\rm diss, r} = \frac{1}{2} \nu_{\rm turb} \left(\frac{\partial \omega}{\partial \ln r}\right)^2
~~({\rm erg~ g^{-1} sec^{-1}})~,
\end{equation}
where $\omega$ is the angular velocity, $r$ the radius and $\nu_{\rm turb}$ the turbulent
viscosity due to the above mentioned rotationally induced instabilities
(Heger et al.~\cite{Heger00a}; Yoon \& Langer~\cite{Yoon04a}).

Nuclear networks for the changes in chemical composition and
the nuclear energy generation include more than 60 reactions 
(see Heger et al.~\cite{Heger00a} for more details).
In particular, the ${\rm ^{14}N(e^{-},\nu)^{14}C(\alpha, \gamma)^{18}O}$
reaction (hereafter NCO reaction), which
becomes active when $\rho \gsim 10^6$~\density{} (Hashimoto et al.~\cite{Hashimoto84}), 
has been newly included for this study. 
We have used the ${\rm ^{14}N(e^{-},\nu)^{14}C}$ reaction rate
given by Martinez (2002, Private communication)
and followed Caughlan and Fowler (\cite{Caughlan88}) 
for the ${\rm ^{14}C(\alpha, \gamma)^{18}O}$
reaction rate.
The accretion induced heating is described following Neo et al. (\cite{Neo77}),
and the accreted matter is assumed to have the same entropy as that of 
the surface of the accreting star.

Two initial masses, 0.6 and 0.8~\Msun{}, are considered for the CO white dwarf models.
Since isolated white dwarfs are generally found to rotate with 
a surface velocity of
$v_{\rm s} \lsim 40$~km/s{}
(Heber et al.~\cite{Heber97}; Koester et al.~\cite{Koester98}; Kawaler~\cite{Kawaler03}),
the initial rotation velocity of our models is assumed to be as slow as 10~km/s at the white dwarf equator 
(see also Langer et al.~\cite{Langer99}).  
Other physical properties of the white dwarf initial models
are summarized in Table~\ref{tab:init}. 
While most of our simulations start with a cold white dwarf 
with $\log L_{\rm s}/{\rm L}_{\odot} \simeq -2.0$,
an initially hot white dwarf with $\log L_{\rm s}/{\rm L}_{\odot} \simeq 2.508$ is also
considered for one model sequence (TC, Table~\ref{tab:results}; cf. Sect.~\ref{sect:henova}).
The accreted matter, received with two different constant accretion rates of 
\Mdot{}~=~\mdota{} and \mdotb{}~\msyr{}, 
is assumed to have $Y=0.982$ and $X_{\rm N} = 0.012$, where $Y$ and $X_{\rm N}$ are
the mass fraction of helium and nitrogen, respectively.

In a close binary system, the white dwarf is believed to
receive matter through a Keplerian accretion disk if its magnetic field
is not considerable.
The accreted matter may thus carry an amount of specific angular momentum
which corresponds to critical rotation at the white dwarf equator.
However, continuous angular momentum gain under these conditions
leads to over-critical rotation soon after
the onset of mass transfer
in the outer part of the accreting star (Yoon \& Langer~\cite{Yoon02}). 
Therefore, we limit the angular momentum gain such that 
the accreting star may not rotate over-critically, 
as follows:
\begin{equation}\label{eq2}
j_{\rm acc} = \left\{ \begin{array}{ll} 
f \cdot j_{\rm{Kepler}}  & \textrm{if $ v_{\rm{s}} < f \cdot v_{\rm{Kepler}}$ } \\
0 &  \textrm{if $ v_{\rm{s}} = f \cdot v_{\rm{Kepler}}$ } \end{array} \right.
\end{equation}
where $v_{\rm s}$ denotes the surface velocity at the white dwarf equator, 
$j_{\rm acc}$ the specific angular momentum of the accreted matter, and
$v_{\rm Kepler}$ and $j_{\rm Kep}$ the Keplerian value 
of the surface velocity and 
the specific angular momentum at the white dwarf equator, respectively.
The dimensionless parameter $f$ represents the fraction of the Keplerian value
of the specific angular momentum which is contained in the accreted
matter, and also the maximum fraction of the critical rotational velocity with
which the white dwarf equator may rotate.

\begin{table}[t]
\begin{center}
\caption{Physical quantities of the initial white dwarf models:
 mass, surface luminosity, central temperature, central density, 
radius and rotation velocity. The hot white dwarf model in the third row 
is only used for sequence TC (see Table~\ref{tab:results}).}\label{tab:init}
\begin{tabular}{c c c c c c }
\hline \hline
$M_{\rm WD, init}$  &  $\log L_{\rm s, init}/{\rm L_\odot}$ & $T_{\rm c, init}$ & $\rho_{\rm c, init}$ & $R_{\rm WD, init}$ & $v_{\rm rot, init} $ \\
    \Msun           &                   &     $10^7$ K      & $10^6~{\rm g/cm^3}$  &  \Rsun             &     km/s \\
\hline
    0.6             &    $-2.049$       &      1.69          &  3.64               & 0.0126            &    10 \\
    0.8             &    $-2.024$       &      1.59          &  10.7               & 0.0101            &    10 \\
    0.8             &     2.508         &      14.4          &  6.44               & 0.0186            &    10 \\
\hline
\end{tabular}
\end{center}
\end{table}

\begin{table*}[t]
\begin{center}
\caption{Properties of the computed model sequences. \Minit{} : initial mass, \Mdot{} : accretion rate, 
$f$ : fraction of the Keplerian value of the accreted specific angular momentum (see Eq.~\ref{eq2}).
$\dot{E}_{\rm diss}$ : Rotational energy dissipation due to frictional heating. Yes (or No) means that this effect
is considered (or not). 
\DMhe{} : accumulated helium mass until the helium ignition point.
$T_{\rm c}$ and $\rho_{\rm c}$ : central temperature and density in the last computed model.
$T_{\rm He}$ : maximum temperature in the helium envelope  in the last computed model.
$\rho_{\rm He}$ and $\eta_{\rm He}$  : density and degeneracy parameter 
at the position of the maximum temperature in the helium envelope in the last computed model.
The last column indicates whether the model sequence will finally result in helium detonation or not.
}\label{tab:results}
\begin{tabular}{c c c c c c c r c c r c}
\hline \hline
No. & $M_{\rm WD, init}$&  \Mdot  & $f$ & $\dot{E}_{\rm diss}$ & $\Delta M_{\rm He}$ & $T_{\rm c}$ & $\rho_{\rm c}$ & $T_{\rm He}$ & $\rho_{\rm He}$ & $\eta_{\rm He}$  & Detonation?  \\
    & \Msun             &  $10^{-8}$ \msyr &     &                      &   \Msun    & $10^8$ K    & $10^6~{\rm g/cm^3}$     & $10^8$ K     & $10^6~{\rm g/cm^3}$  &   \\
\hline
 N1 &   0.6    & 2.0 &  --  & -- & 0.229  & 0.60 & 10.9 & 2.0 & 1.48 & 11 & Yes \\
 N2 &   0.6    & 3.0 &  --  & -- & 0.185  & 0.54 & 8.8  & 1.4 & 0.55 & 9  & No  \\
 N3 &   0.8    & 2.0 &  --  & -- & 0.168  & 0.55 & 26.1 & 2.0 & 1.58 & 13 & Yes \\     
 N4 &   0.8    & 3.0 &  --  & -- & 0.150  & 0.46 & 23.6 & 1.2 & 1.50 & 19 & Yes \\     
\hline
 R1  &   0.6    & 2.0 &  1.0 & Yes & 0.080 & 0.39 & 4.4  & 1.2 & 0.22 & 6  & No \\
 R2  &   0.6    & 3.0 &  1.0 & Yes & 0.061 & 0.23 & 4.2  & 1.4 & 0.15 & 4  & No \\
 R3  &   0.8    & 2.0 &  1.0 & Yes & 0.022 & 0.14 & 11.0 & 1.5 & 0.15 & 4  & No \\
 R4  &   0.8    & 3.0 &  1.0 & Yes & 0.017 & 0.14 & 10.6 & 3.8 & 0.03 & 0.3& No \\
\hline
 TA1  &   0.6    & 2.0 &  0.6 & Yes & 0.149 & 0.50 & 5.97 & 1.2 & 0.42& 9 & No \\
 TA2  &   0.6    & 3.0 &  0.6 & Yes & 0.106 & 0.45 & 4.70 & 3.5 & 0.16& 2 & No \\
 TA3  &   0.8    & 2.0 &  0.6 & Yes & 0.073 & 0.21 & 14.2 & 1.3 & 0.38& 8 & No \\
 TA4  &   0.8    & 3.0 &  0.6 & Yes & 0.045 & 0.18 & 12.7 & 1.4 & 0.28& 6 & No \\
\hline
 TB1  &   0.6    & 2.0 &  1.0 & No  & 0.402 & 0.59 & 9.6 &  1.1 & 1.14 & 22 & Yes \\
 TB3  &   0.8    & 2.0 &  1.0 & No  & 0.233 & 0.65 & 21.3 & 1.2 & 1.36 & 21 & Yes \\
\hline
 TC   &   0.8    & 1.0 &  1.0 &  Yes  & 0.002 & 0.95 &  8.5 & 2.8 & 0.001 & 0.06& No \\
\hline

\end{tabular}
\end{center}
\end{table*}

The use of $f=1$ might be the most natural choice
to describe the realistic situation, as 
Paczy\`nski (\cite{Paczynski91}) and Popham \& Narayan (\cite{Popham91}) 
argue that a critically rotating star may continue to accrete matter
from a Keplerian disk by transporting angular momentum from the star
back into the disk due to turbulent viscosity.
However, as the correct treatment of close-to-critical rotation
is beyond the capabilities of our numerical code, we also consider
the case $f<1$.
As discussed in Yoon \& Langer (\cite{Yoon04a}), 
the correction factors of $f_{\rm P}$ and $f_{\rm T}$ which
are included in the stellar structure equations for describing the effects of
rotation (cf. Heger et al.~\cite{Heger00a}) 
are limited to 0.75 and 0.95, respectively. 
This limit corresponds to a rotation rate of about 60\% of critical
rotation, up to which 
our one dimensional approximation in computing the effective gravitation potential
can accurately describe the structure of the rotating star. 
In our models with $f=1$, where the outer envelope rotates close to
critically, the
centrifugal force is accordingly underestimated.
Although the region which rotates faster than 60\% critical contains
only little mass (see Sect.~\ref{sect:results}), 
we also consider the case $f=0.6$, apart from the case $f=1.0$. 
With $f=0.6$, the white dwarf never rotates faster than 60 \% of the Keplerian value, 
and thus the stellar structure is accurately described throughout the white dwarf interior. 

For comparison,
we also compute rotating models where the rotational energy dissipation is neglected, 
as well as non-rotating models, with otherwise identical initial conditions.
In Table~\ref{tab:results} we list all computed model sequences. 
The index N in the model number indicates a non-rotating model sequence, 
while R denotes a rotating one with $f=1.0$. 
The index TA is for rotating test models with $f=0.6$, while
TB is used for rotating test models with $f=1.0$ without rotational energy 
dissipation (i.e., $\epsilon_{\rm diss,r} = 0.0$). 
The sequence TC designates the only model starting with a hot white
dwarf. All sequences are computed up to the point where the accumulated
helium shell ignites.

\section{Results}\label{sect:results}

We summarize the results of our simulations in Table~\ref{tab:results}. 
Here, $T_{\rm He}$, $\rho_{\rm He}$
and $\eta_{\rm He}$ denote, for the last computed model, 
the maximum temperature in the helium envelope 
and  the corresponding density 
and the degeneracy parameter 
($\psi/kT$, e.g. Kippenhahn \& Weigert 1990).
\DMhe{} gives the accumulated helium mass until helium ignition.

At helium ignition we stop our calculations, and our models are per se
not able to predict whether the helium burning develops into a
helium detonation or not. However, from the literature (e.g., Woosley
\& Weaver~\cite{Woosley94}) we conclude that a helium detonation
can not develop if the ignition density is smaller than
$\sim 10^6$~\density{},
due to the quenching of the thermonuclear runaway by expansion.
For $\rho_{\rm He} < 10^6 $~\density,
therefore, the helium ignition may result
only in a nova-like shell flash, which will not be able to trigger
core carbon ignition.   

Our results for the non-rotating model sequences are found 
in good agreement with similar models computed 
by Woosley \& Weaver (\cite{Woosley94}) and Piersanti et al. (\cite{Piersanti01}). 
For instance, for a sequence with the same initial mass and accretion
rate as for our sequence~N1,
Piersanti et al. obtained \DMhe{}~=~0.244~\Msun{},  
which does not differ much from our result of \DMhe{}~=~0.229~\Msun. 
The slightly smaller value of the present study may be attributed to 
a small difference in the initial nitrogen mass fraction,
which triggers the NCO reaction when $\rho_{\rm c} \gsim 10^6$~\density{}
and initiates helium burning.

\begin{figure}[t]
\resizebox{\hsize}{!}{\includegraphics{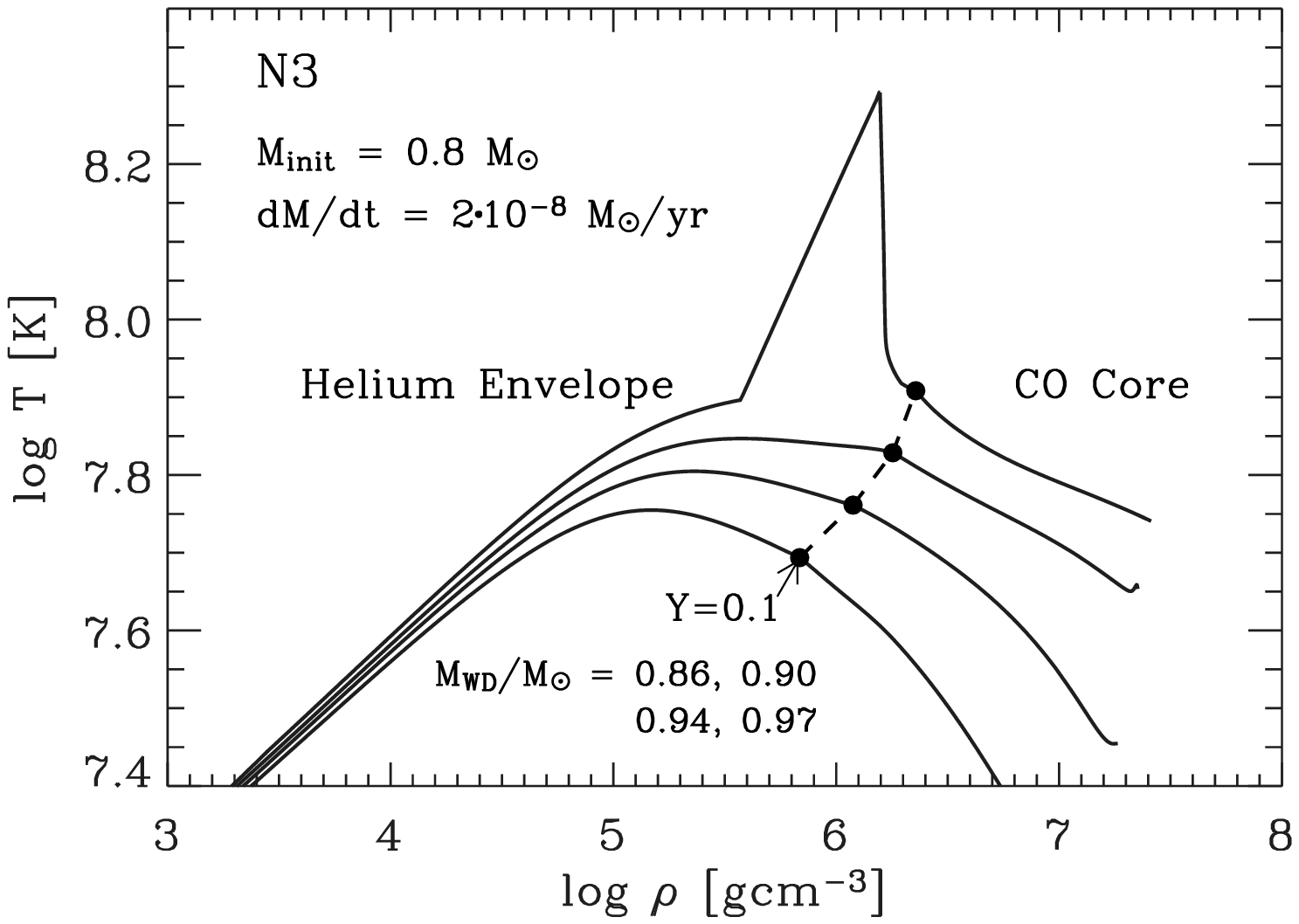}}
\resizebox{\hsize}{!}{\includegraphics{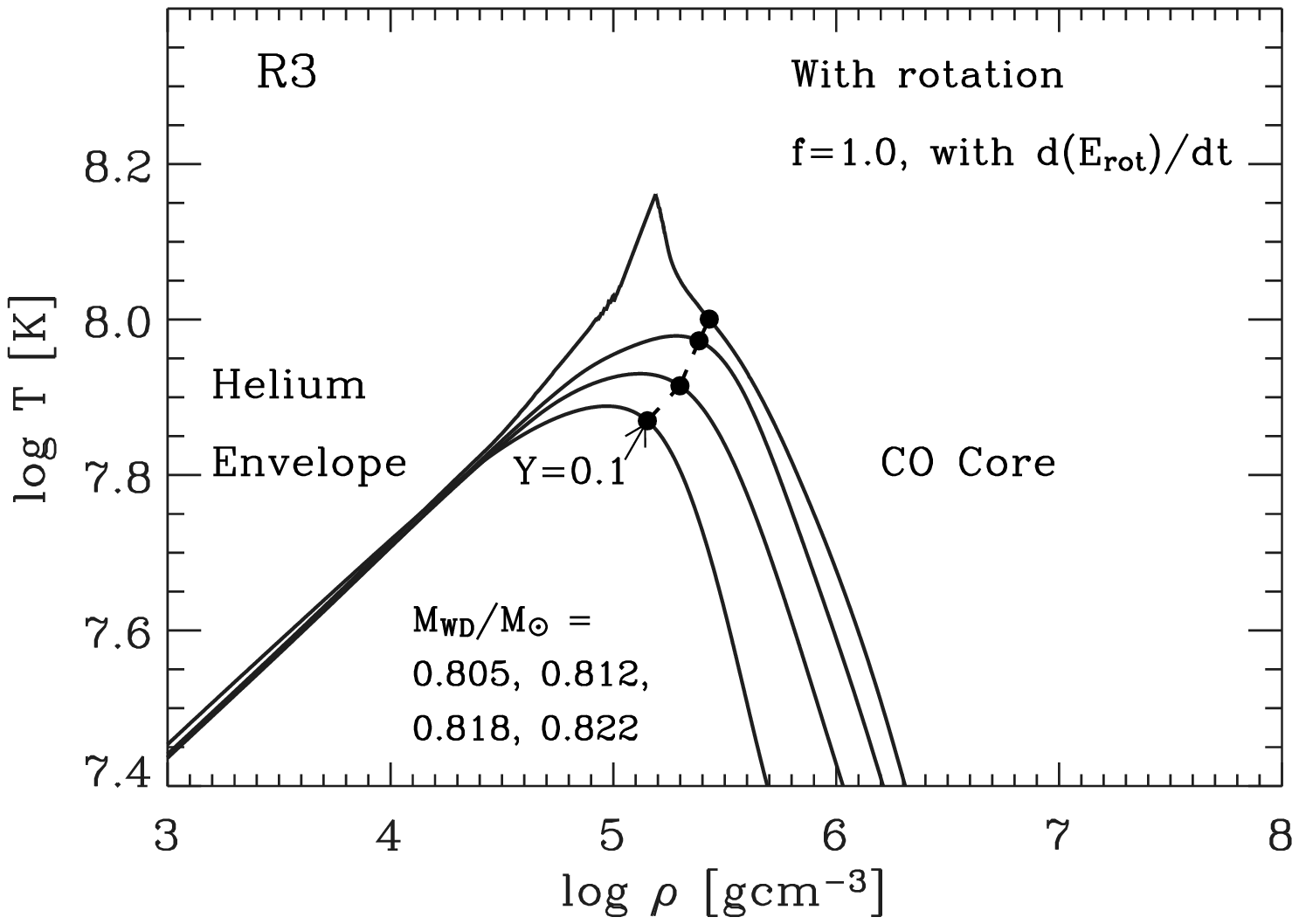}}
\resizebox{\hsize}{!}{\includegraphics{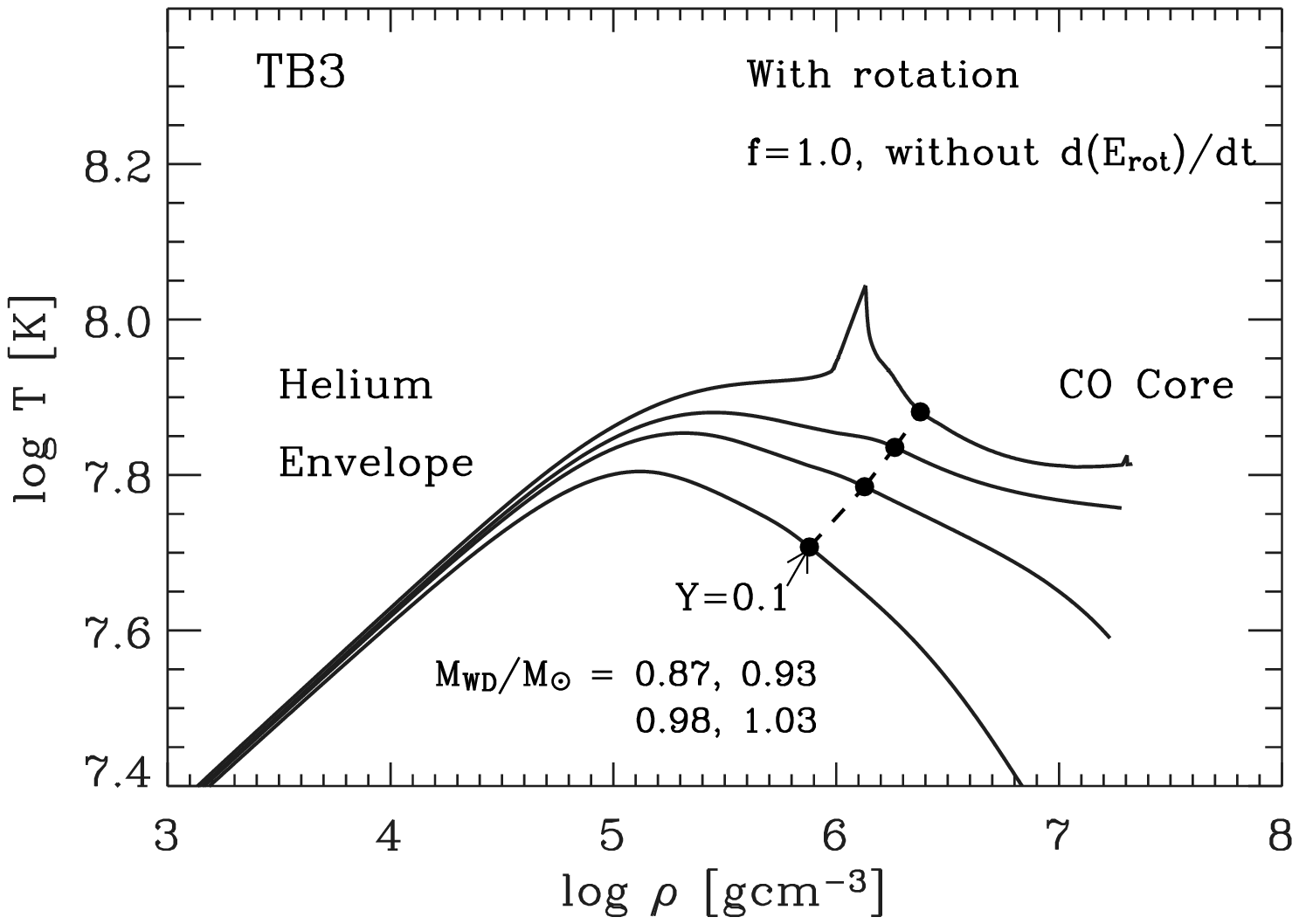}}
\caption{Evolution of accreting white dwarf models of sequence N3, R3 \& TB3,
for which \Minit= 0.8 \Msun{} and $\dot{M}=2\times10^{-8}~{\rm M_{\odot}/yr}$,
in  the $\log \rho - \log T$ plane.
The boundary between the CO core and the helium envelope, defined though
$Y=0.1$, is indicated by a filled circle for each model.
}\label{fig:evol}
\end{figure}

Fig.~\ref{fig:evol} illustrates the evolution of our white dwarf models 
with \Minit{} = 0.8~\Msun{}
and \Mdot{} = \mdota{}~\msyr{} (sequences N3, R3 and TB3). 
In the non-rotating case, the temperature 
of the helium envelope continues to increase due to accretion induced heating.  
When the white dwarf mass reaches about 0.9~\Msun, 
the density at the bottom
of the helium envelope starts exceeding $10^6$~\density{}
and the NCO reaction becomes active, 
accelerating the temperature increase.
Finally, helium burning starts when \Mwd{}  reaches 0.968~\Msun. 
The density at the helium ignition point is about $1.6 \times 10^6$~\density{}
and a detonation is likely to follow.

The evolution of the corresponding rotating model (R3) in
Fig.~\ref{fig:evol} looks similar, but 
helium ignites much earlier than in the non-rotating case.
The maximum temperature in the helium envelope
reaches $10^8$ K  when \Mwd{}~$\simeq$~0.818~\Msun. 
Helium burning develops quickly thereafter, and the nuclear energy
generation amounts to $\log L_{\rm He}/L_{\odot} \simeq  5.0 $ 
when  \Mwd{}~$\simeq$~0.822~\Msun. Only 0.022~\Msun{} of helium has
been accumulated by then, which is about 10 times less than
in the non-rotating case. Importantly, the ignition density
$\rho_{\rm He}$ is also found about 10 times lower ($1.5 \times 10^5$~\density{})
than in the non-rotating case, implying 
that a helium detonation may be avoided. 

\begin{figure}[!]
\center
\resizebox{\hsize}{!}{\includegraphics{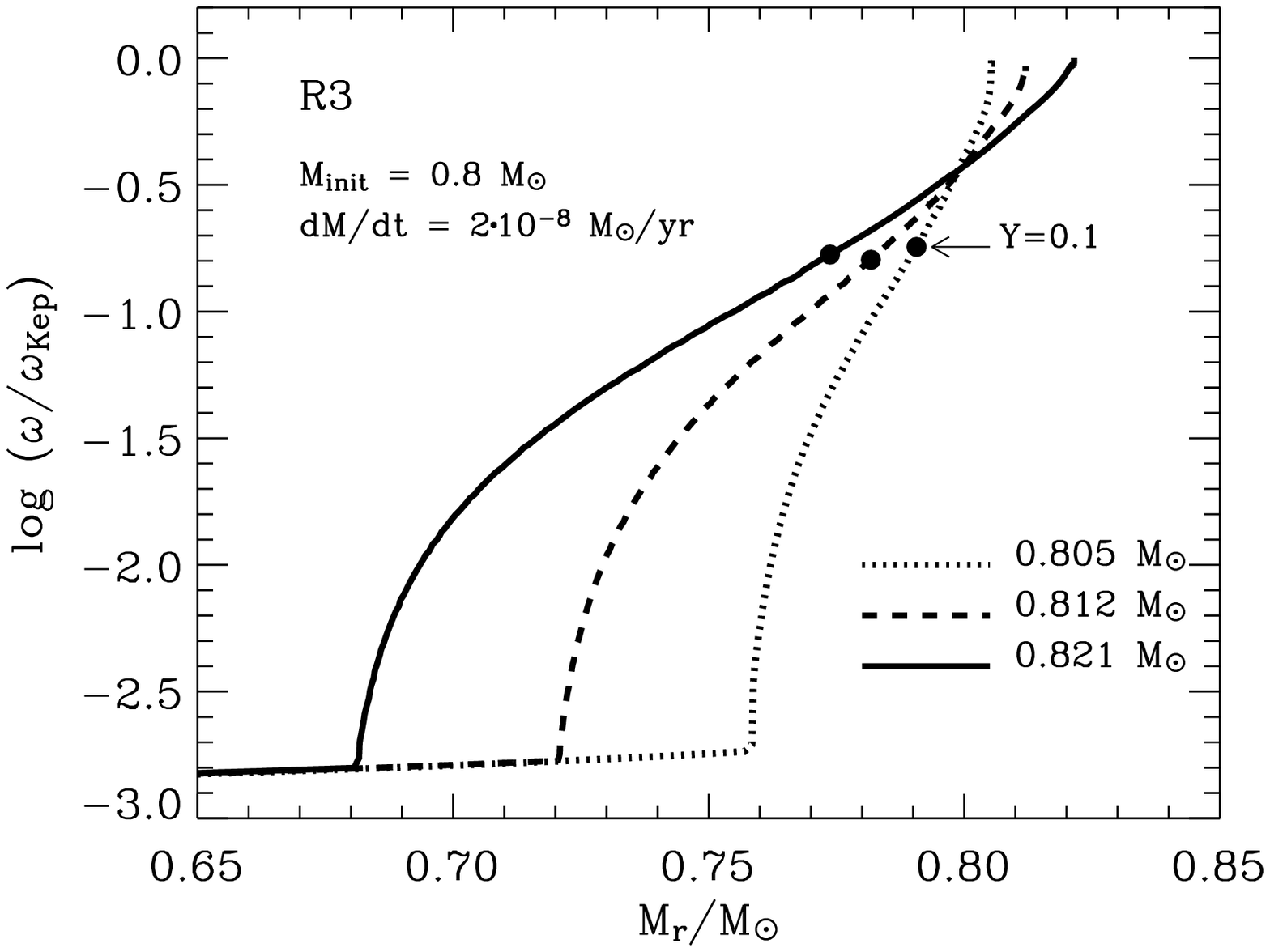}}
\resizebox{\hsize}{!}{\includegraphics{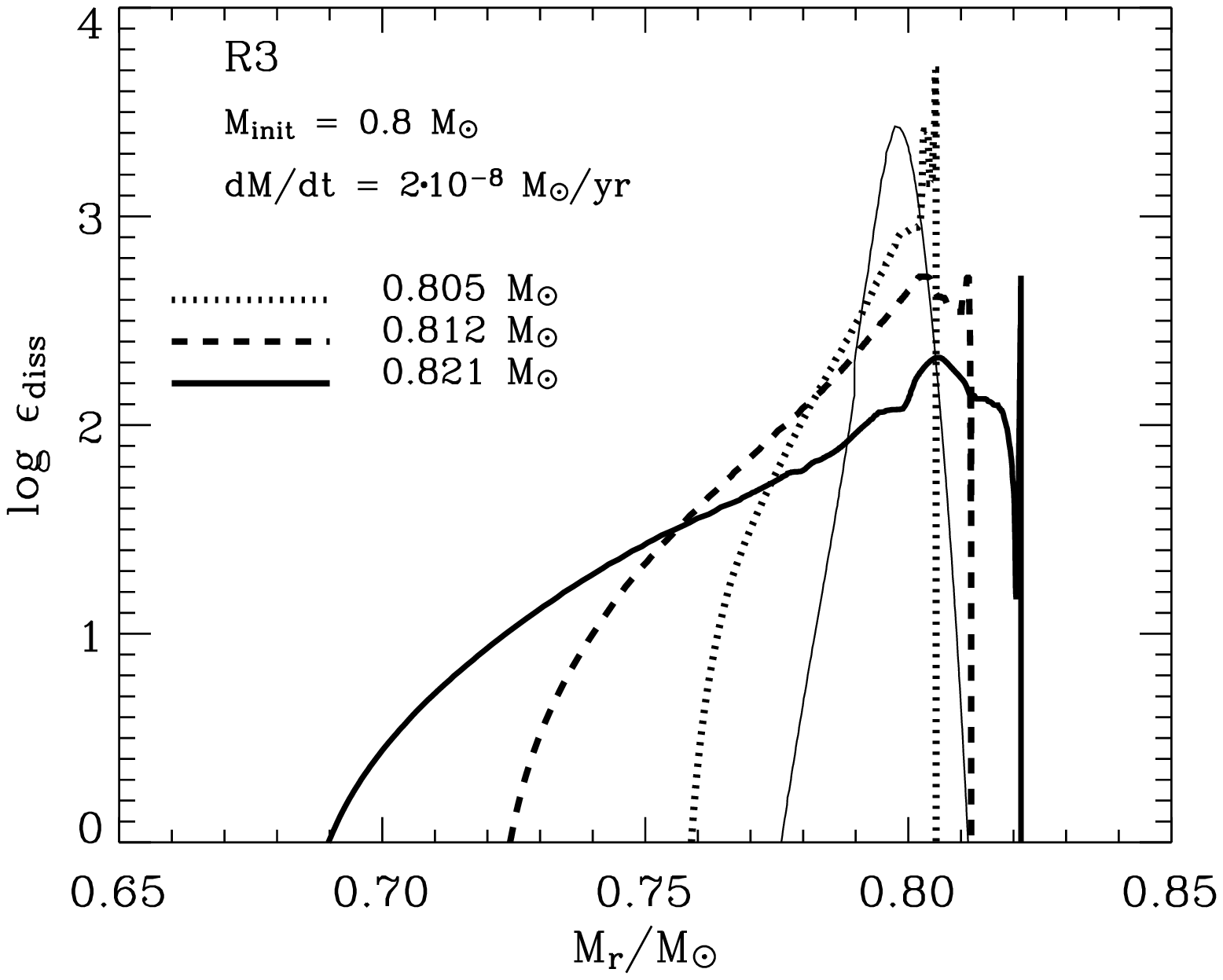}}
\caption{{\em Upper panel}: Angular velocity in units of the local Keplerian value as function
of the mass coordinate in model sequence R3 when \Mwd{}~= 0.805, 0.812 and 0.821~\Msun.
The filled circles indicate the boundary of the CO core and helium envelope,
as defined in Fig.~\ref{fig:evol}, where the helium mass fraction is 0.1.
{\em Lower panel} : Rotational energy dissipation due to friction (Eq.~\ref{eq1})
at the same evolutionary epochs as in the upper panel.
The thin solid line denotes the energy generation rate due to nuclear reactions
when \Mwd{} = 0.821~\Msun.
}\label{fig:spin}
\end{figure}

The reason for the early helium ignition in the rotating sequence is as follows.
As explained above, the white dwarf obtains angular momentum
carried by the accreted matter. The angular momentum is thus
transported from the surface into the interior by various
rotationally induced hydrodynamic instabilities. 
In our white dwarf models, Eddington Sweet circulation, 
the secular shear and the GSF instability
dominate in the non-degenerate envelope, while
the dynamical shear instability is most important
in the degenerate interior, as 
discussed in detail by Yoon \& Langer (\cite{Yoon04a}).
The white dwarf interior is spun up progressively with time, 
giving rise to rotational energy dissipation
as shown in Fig.~\ref{fig:spin}. 
In the upper panel of the figure,
the angular velocity
reveals a strong degree of differential rotation. 
The consequent energy dissipation rate 
is given in the lower panel in the same figure.
The total energy dissipation rate integrated 
over the spun-up layers in the given models
amounts to 8.8~\Lsun, 6.2~\Lsun, and 3.6~\Lsun{} 
for \Mwd{}~=~0.805~\Msun, 0.812~\Msun, and 0.821~\Msun{}
respectively. 
Together with the accretion induced heating,  
this additional energy supply speeds up the temperature increase
in the helium envelope, leading to the earlier helium ignition
compared to the non-rotating case. 

For comparison, the white dwarf evolution in sequence TB3, 
where the rotational energy dissipation is neglected while
everything else is as in sequence R3,
is shown in the third panel of Fig.~\ref{fig:evol}. In this case, 
helium ignition occurs even later than in
the corresponding non-rotating case (N3).
This is due to the lifting effect of the centrifugal force, 
which reduces the accretion induced heating.
The density at the bottom of the helium envelope
when the helium burning is induced by the NCO reaction
is about  $1.36 \times 10^6$~\density{}, implying
that a helium detonation might be triggered in this case.

As shown in Fig.~\ref{fig:spin},
the surface layers of the white dwarf models in sequence R3, where $f=1$ is used, 
rotate faster than 60 \% critical, above which
the effects of rotation are underestimated by our computational method.
In the models with \Minit{} = 0.8~\Msun{}, i.e., in sequences R3 and R4, however, 
this fast rotating region is limited to only 1\% of the white dwarf mass, 
which is not likely to affect the result significantly. 
On the other hand, about 7\% in mass, which include a significant fraction of the helium envelope,
exceed the limit at helium ignition
in model sequences R1 and R2, for which \Minit{} = 0.6~\Msun{} and $f=1$ are adopted.
Nevertheless, results from the corresponding sequences with  $f=0.6$ (TA1, TA2, TA3 and TA4), 
where the white dwarf models are forced to rotate below 60\% critical throughout
the white dwarf interior, lead us to the same conclusion as in case of $f=1$: 
helium ignites at such a low density (even though it is somewhat 
higher than in case of $f=1$) that a supernova event is unlikely to occur.
For instance, in model sequence TA1, we have $\rho_{\rm He} = 4.2 \times 10^5$~\density{}
when the maximum temperature reaches $1.2 \times 10^8$ K. 
This is about 3.5 times lower than in the corresponding non-rotating case 
($\rho_{\rm He} = 1.48 \times 10^6$~\density)
and about two times higher than in case of $f=1$ ($\rho_{\rm He} = 2.2 \times 10^5$~\density).


\section{Discussion}\label{sect:discussion}

\subsection{Connection to recurrent helium novae}\label{sect:henova}

As discussed by Iben \& Tutukov (\cite{Iben91}) and Limongi \& Tornamb\'e (\cite{Limongi91}), 
a relatively low helium accretion rate of $\sim 10^{-8}$~\msyr{} 
onto a CO white dwarf can be realized, for example, in
binary systems which consist of a 0.6~$\cdots$~1.0~\Msun{} CO white dwarf
and  a less massive helium star. Mass transfer in such binary systems
is driven by gravitational wave radiation, and the resulting
mass transfer rate is found to be insensitive to the exact
masses of white dwarf and helium star, at a few times $10^{-8}$~\msyr{} 
(Iben \& Tutukov~\cite{Iben91}; Limongi \& Tornamb\'e~\cite{Limongi91}). 
Many of such binary systems, if not all,  could produce a supernova
which will appear as Type Ia 
(Taam~\cite{Taam80}; Nomoto~\cite{Nomoto82b}; 
Iben \& Tutukov~\cite{Iben91}; Limongi \& Tornamb\'e~\cite{Limongi91}; 
Woosley \& Weaver~\cite{Woosley94}; Livne \& Arnett~\cite{Livne95}), 
while the resulting light curves and spectra 
may be abnormal 
(H\"oflich \& Tutukov~\cite{Hoeflich96}; Nugent et al.~\cite{Nugent97}). 
As mentioned in the introduction of this paper, binary
population synthesis models show that the production rate of such events may be comparable 
to the observed SN~Ia rate (e.g. Iben \& Tutukov~\cite{Iben91}; Reg\"os et al.~\cite{Regos02}), 
and therefore the apparent absence of such a
peculiar type of supernovae has been a puzzling matter.

The results of the present study offer a possible solution to this problem:
helium accreting CO white dwarfs with an accretion rate of $\sim 10^{-8}$~\msyr{}
may not result in a supernova at all, 
but may instead produce nova-like explosions.
The accumulated helium mass of 
\DMhe{}~$\simeq 0.02 \cdots 0.1$~\Msun{} 
in our rotating models indicates
that in a low mass helium star + CO white dwarf binary system, 
nova explosions will occur recurrently, with a period of about $10^6$ yr. 

As pointed out by Iben \& Tutukov (\cite{Iben91}), various
factors need to be considered for predicting
the further evolution of such a binary system
after the first nova outburst.
First, any mass loss induced by the helium flash  
will affect the binary orbit. I.e., 
the nova induced mass loss will widen the orbit and interrupt the mass
transfer for some time (Iben \& Tutukov~\cite{Iben91}).
However, the angular momentum loss due to gravitational wave radiation 
will lead the helium star to fill its Roche lobe soon again (after $\sim 10^{6}$ yr) 
as discussed below.

\begin{figure}[t]
\center
\resizebox{\hsize}{!}{\includegraphics{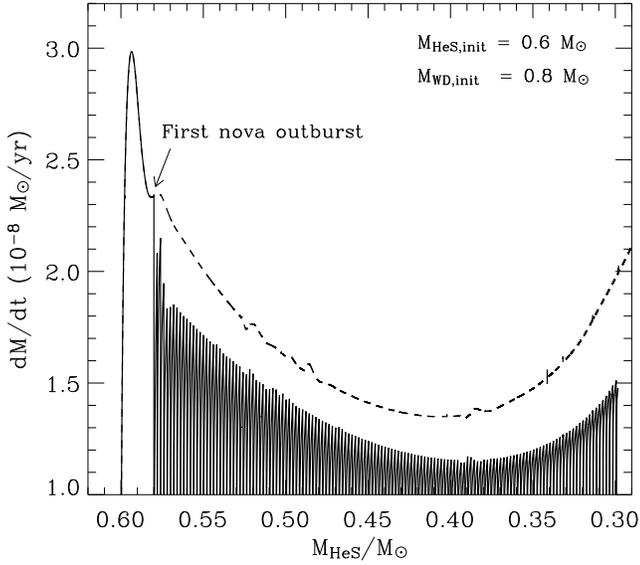}}
\caption{Mass transfer rate as function of the helium star mass, for
a binary system consisting initially of an 0.6 \Msun{} helium star and an 0.8 \Msun{} CO white dwarf (the latter here being approximated by a point
mass). 
Mass transfer is initiated by gravitational wave radiation.
The solid line illustrates the case where mass loss due to helium flashes
is considered,
while the dashed line was computed assuming that no mass loss occurs.
}\label{fig:mdot}
\end{figure}

Second, the first nova outburst may
heat up the white dwarf significantly,
and the white dwarf may still be hot when the second
mass transfer starts, compared to the case of
the first mass transfer.
Model sequence TC, where an initially hot white dwarf with $\log L_{\rm s}/{\rm L}_{\odot} = 2.508$ 
is adopted, was computed in order to investigate 
the effect of a pre-heated white dwarf.
A value of \Mdot{} = $10^{-8}$~\msyr{}
for this sequence, which is smaller than in other model sequences,
was chosen to consider the decrease of the mass transfer rate due to the change of 
the orbit as the binary system loses
mass via helium nova outbursts (cf. Fig~\ref{fig:mdot}).
A comparison of \DMhe{} in sequence TC (\DMhe{} $\simeq$ 0.002~\Msun) 
with the accumulated helium masses in the sequences 
R3 and R4 (\DMhe{} $\simeq$ 0.02~\Msun)
implies that \DMhe{} decreases by more than a factor of 10 
if the white dwarf is preheated.
Therefore, the second and any further helium flash may be much weaker
than the first one.

\begin{figure}[t]
\center
\resizebox{\hsize}{!}{\includegraphics{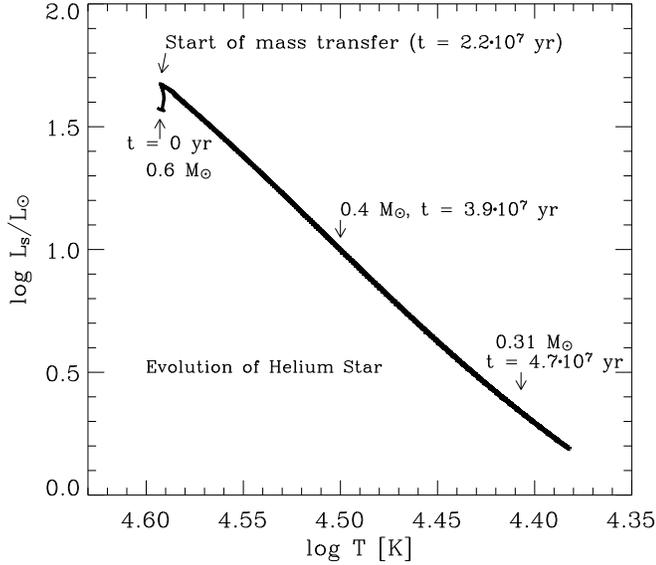}}
\caption{
Evolution in the HR diagram of the helium star 
(initial mass 0.6 \Msun{}) component of the binary system considered in
in Fig.~\ref{fig:mdot}, for case with nova induced mass loss.
}\label{fig:hr}
\end{figure}

In an attempt to make more detailed conjectures 
on the evolution and final fate of binary systems
of the considered kind,
we made a simple experiment as follows.
We constructed a binary star model
consisting of a zero-age helium main sequence star of 0.6~\Msun{}
and a CO white dwarf of 0.8~\Msun{} in an 1.08~h orbit 
($A_{\rm orbit, init} = 0.6$~\Rsun{}, cf. Limongi \& Tornamb\'e~\cite{Limongi91}).
Here, the white dwarf is approximated by a point mass.
Since it appears likely that the accreted helium 
will be ejected from the system by the violent helium shell flash
when about 0.02 \Msun{} of helium is accumulated as implied by 
the results of sequences R3 and R4,
we assume this for the evolution of our binary system.
As discussed above, 
the subsequent helium flashes may occur with a smaller \DMhe{}.
Therefore, for the subsequent evolution 
the white dwarf is assumed to lose the accumulated matter
at every time when \DMhe{} = 0.002~\Msun{} is achieved.
For comparison, a second binary evolution model with the same initial condition 
is calculated, with the assumption of no mass loss due to helium flashes.
The evolution of the helium star and
the change of the binary orbit due to the mass transfer, stellar wind mass loss and
gravitational wave radiation are followed by using a
binary stellar evolution code
(see Langer et al.~\cite{Langer00} for more details 
about the code).

Fig.~\ref{fig:mdot} shows the evolution of the mass transfer rate in the considered system, 
as function of the helium star mass (\Mhes).
The mass transfer from the helium star starts when the helium 
mass fraction in the helium star center equals 0.39. 
The orbital period at this moment is about 39~min.
The mass transfer rate initially increases to $3\times10^{-8}$~\msyr{}
and has decreased to $2.3 \times 10^{-8}$~\msyr{} by the time \DMhe{} reaches 0.02~\Msun. 
As the white dwarf is assumed to lose 0.02~\Msun{} at this point due to the helium nova flash,
the system becomes detached for about $5.5 \times 10^5$ yr, after
which the helium star again fills its Roche lobe. 
Further-on, the white dwarf loses
mass whenever \DMhe{} reaches 0.002~\Msun{}, and thus
the mass transfer is switched on and off repeatedly every $\sim 1.5 \times 10^5$ yr.
We follow the evolution of the system until the helium star mass decreases to 0.30~\Msun{}. 

Fig.~\ref{fig:hr} shows the evolution of the helium star in the HR diagram. 
The helium star luminosity decreases continuously as it loses mass, and 
it will finally evolve into a white dwarf. 
At this stage, the binary system will resemble an AM CVn system. 

At the end of the calculation (i.e., \Mhes{} = 0.30~\Msun),  the central helium abundance has
decreased to 0.16, and more than 0.1~\Msun{} of helium is still available in the envelope
for further mass transfer.
Since about 135 nova outbursts occurred until the end of the calculation, 
more than 180 recurrent nova outbursts in total are expected to 
occur throughout the evolution of the considered binary system. 
This implies that helium nova explosions in low mass helium star + CO white dwarf 
binary systems could be realized with a frequency of $\sim 0.1~{\rm yr^{-1}}$
in our Galaxy, 
given that such binary systems are being produced at a frequency 
of $\sim 0.001~{\rm yr^{-1}}$ (Iben \& Tutukov~\cite{Iben91}; Reg\"os et al.~\cite{Regos02}).
The recently discovered outburst of V445 Puppis, which has been attributed
to a helium nova (Ashok \& Banerjee~\cite{Ashok03}; Kato \& Hachisu~\cite{Kato03}),
may be a promising observational counterpart of such an event.
The high observed carbon abundance in this system (Ashok \& Banerjee~\cite{Ashok03})
might be explained by the rotationally induced
chemical mixing in the accreting white dwarf, the effects of which are discussed in the next section.

\subsection{Rotationally induced chemical mixing and neutron capture nucleosynthesis}

\begin{figure}
\center
\resizebox{\hsize}{!}{\includegraphics{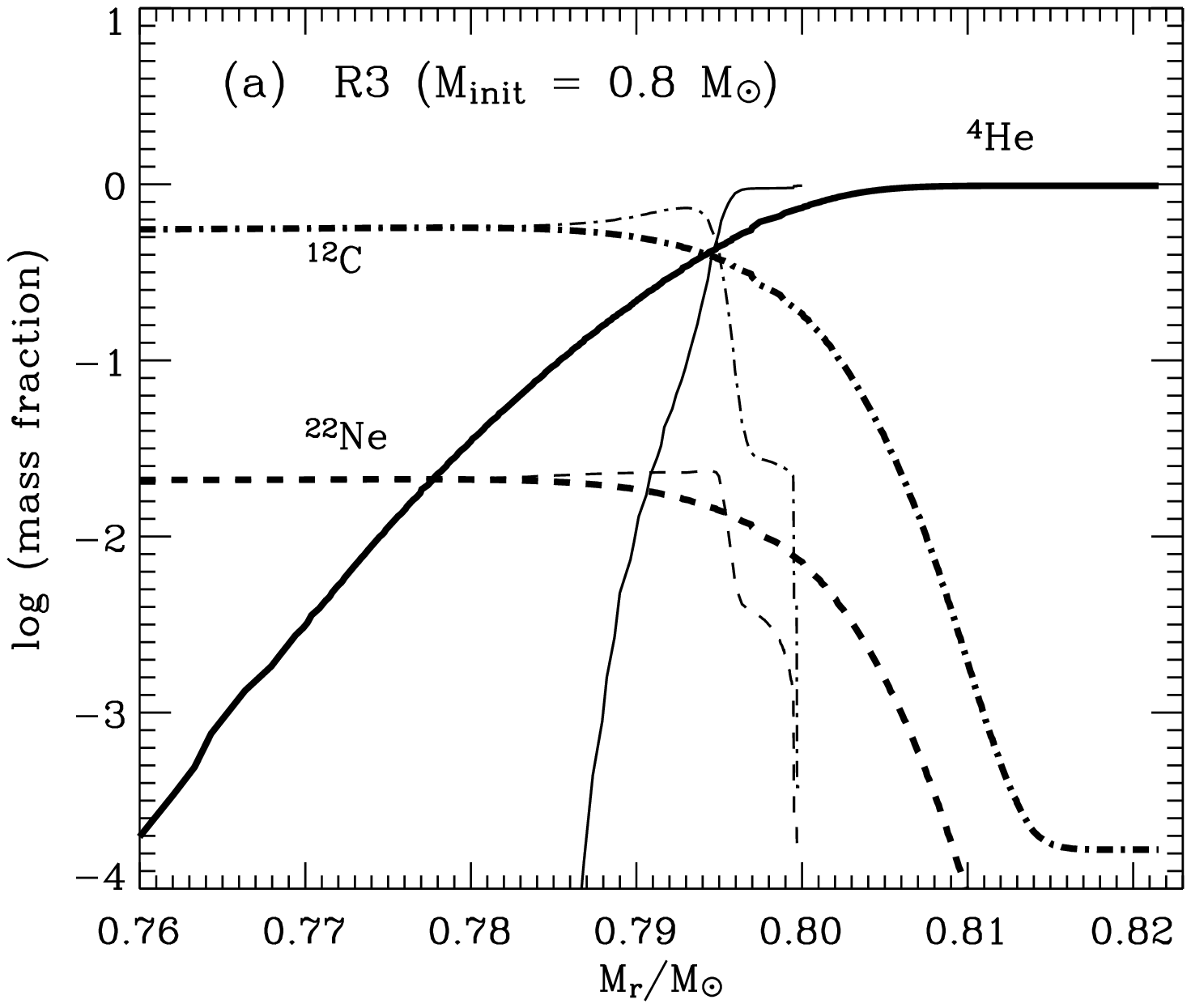}}
\resizebox{\hsize}{!}{\includegraphics{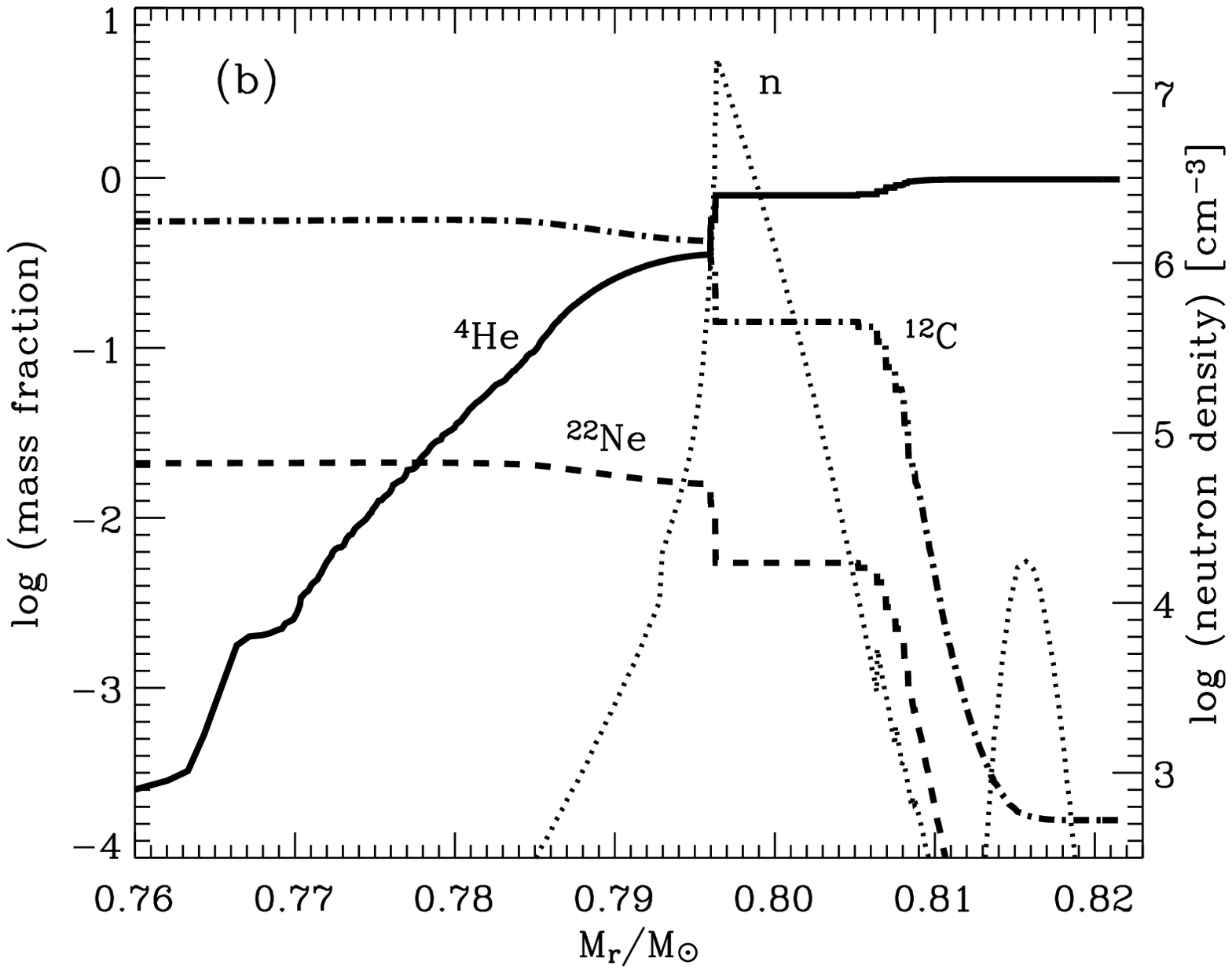}}
\caption{(a) Chemical profiles in the outer layers of the white dwarf models of sequence R3
when \Mwd{} = 0.8~\Msun{} (initial model, thin lines), and \Mwd{} = 0.822~\Msun{} (thick lines)
which is immediate before the helium ignition.
The solid and dotted-dashed lines give the mass fraction of \He{4} and \C{12} respectively, 
as function of the mass coordinate. 
The dashed line denotes the mass fraction of \Ne{22}. 
(b) Same as in (a) but after the ignition of helium. The dotted
line denotes the neutron density, for which the scale is given 
at the right side. 
}\label{fig:mixing}
\end{figure}

As already suggested by Iben \& Tutukov (\cite{Iben91}), 
strong helium flashes in accreting CO white dwarfs may activate neutron capture nucleosynthesis,
since neutrons can be provided via the $^{22}\mathrm{Ne}(\alpha, n)^{25}\mathrm{Mg}$ reaction.
We note that favorable conditions for neutron capture
nucleosynthesis might be
achieved with rotation compared to the non-rotating case, 
due to the rotationally induced chemical mixing (cf. Langer et al.~\cite{Langer99}). 

Fig.~\ref{fig:mixing}a gives the mass fraction of \He{4}, \C{12} and  \Ne{22}
in the initial model (\Mwd{} = 0.8~\Msun) of sequence R3, 
as well as in the model with \Mwd{} = 0.822~\Msun{},
which is immediate before the helium ignition.
It is found that \C{12} and \Ne{22} have been significantly dredged up into the helium envelope 
when helium ignites, due to rotationally induced mixing.
For instance, comparing the two models shown in Fig.~\ref{fig:mixing}a reveals
that the mass fraction of \C{12} and \Ne{22} at $M_{\rm r} = 0.8$~\Msun{}
has increased by about a factor of 10 when helium ignites.

The ignition of helium induces
convection. Fig.~\ref{fig:mixing}b shows the chemical structure 
when $T_{\rm He}$ reaches $1.5 \times 10^8$ K.  
At this point, the convective layer has a mass of $\Delta M_{\rm conv} \simeq 0.01$~\Msun, 
extending from $M_{\rm r} = 0.796$~\Msun{} to $M_{\rm r} = 0.806$~\Msun{}.
The abundance of the dredged up core material in the convective layer 
amounts to 0.15 for \C{12} and $5.5\times10^{-3}$ for  \Ne{22}, respectively. 
Neutrons are released by the $^{22}\mathrm{Ne}(\alpha, n)^{25}\mathrm{Mg}$ reaction, 
and a 
maximum neutron density of $\sim 10^7~\mathrm{cm^{-3}}$ is achieved at 
$M_{\rm r} = 0.76$~\Msun{} in our last model.
This neutron density may be somewhat overestimated in our calculation, since
we did not include the reaction $^{14}\mathrm{N}(n,p)^{14}\mathrm{C}$ in our network 
(cf Siess, Goriely \& Langer~\cite{Siess03}). 
However, we note that alpha particles are mixed significantly into the \Ne{22}-rich
region, and that large amount of neutrons will be released in the further
evolution of the flash which is not covered any more by our model, 
as hydrodynamic effects in the exploding layers may become important.
Therefore, we leave the exploration of the neutron capture nucleosynthesis
effects in our models for the future.

\section{Concluding remarks}\label{sect:conclusion}

We have shown that the effects of rotation in helium accreting white dwarfs 
may be incompatible with the scenario
of double detonations in sub-Chandrasekhar mass CO white dwarfs
as possible SNe Ia progenitors. 
In helium accreting white dwarfs, we find the thermal evolution to be affected  
by viscous heating due to differential rotation in the spun-up
layers, such that helium ignition is induced
at too low densities to develop a detonation. 
This may give a plausible solution to the long standing problem
of the missing observational counterparts of 
sub-Chandrasekhar explosions, which are predicted to occur 
with a frequency comparable to 
the observed SN Ia rate. 

We discussed that binary systems consisting of a CO white dwarf and
a less massive helium star may be possible progenitors of recurrent helium novae 
(Iben \& Tutukov~\cite{Iben91}), 
which may be analogous to V445 Puppis 
(Ashok \& Banerjee~\cite{Ashok03}; Kato \& Hachisu~\cite{Kato03}). 
After the first strong helium nova flash in such binary systems,
rather mild nova outbursts are expected to occur recurrently with a period of $\sim 10^5$ yr.
The realization frequency of such a helium nova may be as high as $\sim 0.1~{\rm yr^{-1}}$
in our Galaxy. 

Rotation induces chemical mixing of \Ne{22} and \He{4}
at the bottom of the helium envelope, which may provide interesting
conditions for neutron capture nucleosynthesis
to occur during the helium nova flashes.

Finally, we note that other important mechanisms 
for the angular momentum redistribution in white dwarfs may exist
than those considered in the present study.
In particular, 
we neglected the possible role of magnetic fields, which
may increase the efficiency of the angular momentum transport
significantly (cf. Heger et al.~\cite{Heger03}; Maeder \& Meynet~\cite{Maeder03}).  
If the spin-up time scale is shorter than considered in the present study, 
the resulting shear strength will be weaker and the effect of rotational energy 
dissipation may not be as important as shown here. 
However, studies of magnetic effects in accreting white dwarfs 
have to be left for future investigations.

\begin{acknowledgements}
We are very grateful to Gabriel Martinez for kindly providing us with 
a data table of the ${\rm ^{14}N(e^{-},\nu)^{14}C}$ reaction rate. 
We thank Onno Pols for fruitful discussion.
This research has been supported in part by the Netherlands Organization for
Scientific Research (NWO).
\end{acknowledgements}

\end{document}